\documentclass[aps,a4paper,preprint,longbibliography]{revtex4}%
\usepackage{amsfonts}
\usepackage{amsmath}
\usepackage{amssymb}
\usepackage{graphicx}
\usepackage{setspace}
\usepackage{bm}%
\providecommand{\U}[1]{\protect\rule{.1in}{.1in}}
\begin{document}
\title{Viscous friction acting on a solid disk falling in confined fluid: lessons for
the scaling analysis}
\author{Nana Tanaka and Ko Okumura}
\affiliation{Physics Department and Soft Matter Center, Ochanomizu University, 2-1-1
Ohtsuka, Bunkyo-ku, Tokyo 112-8610, Japan}
\date{\today}

\begin{abstract}
We fill a viscous liquid in a vertically stood cell of millimeter thickness,
called the Hele-Shaw cell, and insert a disk in the liquid whose thickness is
smaller than the cell thickness. The disk starts falling in the liquid due to
gravity with opposed by viscous friction. We focus on the case in which
lubricating films formed in the gap between the cell surface and the disk
surface are thinner than the disk thickness. As a result, we find an apparent
scaling regime for the falling velocity of a disk, in which the thickness of
the lubricating film characterizes the dynamics. We further show that the
apparent scaling regime is explained simply as a result of competition of two
scaling regimes, elucidating physics of the viscous friction to make the
present study relevant to fundamental issues and applications in various
fields such as microfluidics, bioconvection, and active matter. The simple
scenario for explaining an apparent scaling law demonstrated in the present
study would be useful in diverse fields.

\end{abstract}
\maketitle
\affiliation{Physics Department and Soft Matter Center, Ochanomizu University, Japan}
\affiliation{}

Scaling analysis has been a powerful tool in various fields in science beyond
physics \cite{de1979scaling,Cardy} and applied mathematics
\cite{barenblatt2003scaling}, which include biology
\cite{bonner1983size,schmidt1984scaling}, engineering
\cite{gibson2010cellular}, and astronomy \cite{choudhuri2010astrophysics}. A
lot of scaling laws have been established with a well defined scaling
exponent, with successful explanations. However, it also happens frequently
that, although an exponent is clearly observed, a simple scaling argument (or
an appropriate reorganization group analysis \cite{de1979scaling,Cardy})
predicts an exponent slightly deviating from the observed one, as in, for
example, the metabolic-rate ($R$) vs body-mass ($M$) relation in biology
\cite{kleiber1932body} and the luminosity ($L$) vs mass ($M$) relation in
astronomy \cite{boehm1989introduction}: in the former (latter) experimentally
$R\simeq M^{\alpha}$ ($L\simeq M^{\beta}$) with $\alpha\approx0.75$
($\beta\approx3.7$) while a simple dimensional analysis gives $\alpha=5/8$
($\beta=3$). Such apparent scaling laws have been interpreted in case-specific
and complex theories, leaving room for debate. Here, we focus on a simple
viscous dynamics and demonstrate an apparent scaling law, which is simply
explained as a result of competition of two scaling regimes. This simple
scenario would be useful for understanding apparent scaling laws in diverse
fields in science.

In addition to the significance in the general context of the scaling
analysis, the present study could be fundamentally important for various
current topics such as microfluidics \cite{SquiresQuake2005}, promising for
applications in chemistry, biology, medicine and pharmaceutical industry, and
bioconvection \cite{bees2020advances,kage2013drastic,ramaswamy2010mechanics},
which is created via swimming of a large number of small objects and has
received attention in active matter and biology. The central issue is the flow
at low Reynolds numbers governed by viscous friction due to small scales at
which most liquids including water are highly viscous.

Relevant classic studies on viscous friction are motion of a bubble in a
capillary tube \cite{Bretherton} and in a Hele-Shaw cell
\cite{TAYLORSAFFMAN1959}. Our focus is on the latter. A number of studies have
been performed in the Hele-Shaw geometry, with a main focus on a forced flow
and/or with nearly horizontal geometries
\cite{Tanveer1986,Maxworthy1986,Kopf-SillHomsy1988,MaruvadaPark1996}.
Recently, we have highlighted the existence of lubricating film between the
cell surface and bubble surface, using a vertically stood Hele-Shaw cell of
millimeter thickness, filled with viscous liquid to confirm a number of
scaling regimes for drag friction acting on fluids surrounded by another
immiscible fluid
\cite{EriSoftMat2011,yahashi2016,okumura2017AdvCI,murano2020rising}. Other
groups have explored closely related issues, using cells with smaller scales
\cite{keiser2018dynamics,keiser2019motion} and comparing numerical results
with experiment \cite{shukla2019film}.

In this study, we explore a seemingly simpler case of drag friction acting on
a solid disk in a Hele-Shaw cell of millimeter thickness, where similar
lubricating films exist between the cell wall and disk surface. Our principal
interest is in the case in which the thickness of the lubrication film is
smaller than the disk thickness when the disk falling in the direction
perpendicular to its axis. As far as we know, this case has not been explored
in the literature, while it is closely related a study of transport of
strongly confined disks under flow in microfluidic devices
\cite{uspal2013engineering}. As a result, we identified an apparent scaling
regime, in which the observed scaling exponent for collapsed data deviates
from a predicted exponent. We reveal that the deviation is simply the result
of competition of two scaling regimes.

\begin{figure}[h]
\includegraphics[width=0.9\textwidth]{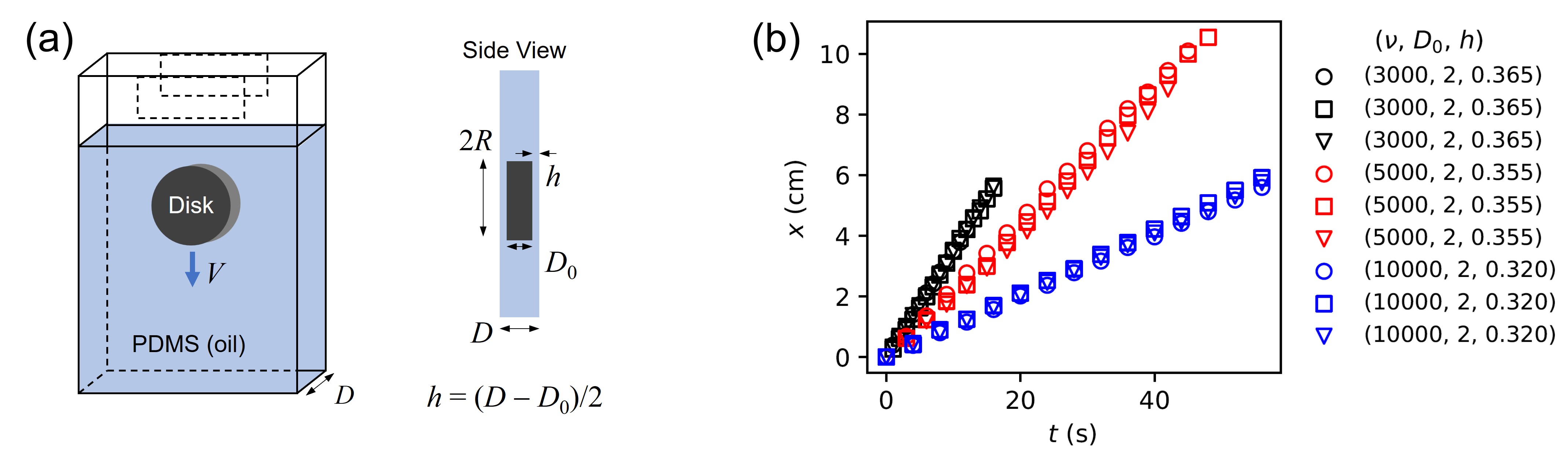}\caption{(a) Illustration of
experiment. A metal disk of thickness $D_{0}$ is dropped in the cell of width
$D$ filled with a viscous oil of kinematic viscosity $\nu$. The side view
suggests the existence of two lubricating films of thickness $h$. The two
dashed rectangles at the top of the cell on the left indicate the places at
which plates (of thickness $\simeq h$) are set (see the text for the details).
(b) Vertical position $x$ vs time $t$. Results of three falling experiments
are shown for each parameter set specified by $(\nu,D_{0},h)$. $\nu$ is given
in cSt, while $D_{0}$ and $h$ are given in mm. (See the text for the precise
value $D_{0}$)}%
\label{Fig1}%
\end{figure}

In experiments, we filled a Hele-Shaw cell, stood vertically, with a silicone
oil (polydimethylsiloxane, PDMS) as shown in Fig. 1 (a). We inserted a metal
disk of radius $R=10$ mm at the top of the cell with a zero initial speed. We
recorded the ensuing falling motion of the metal with a camera after the disk
began going down in the cell at a constant speed. The width and height of the
cell was 90 and 160 mm, respectively. We checked the thickness $D$ of the cell
using a laser sensor (ZS-HLDS5, Omron) and its controller (ZS-HLDC11, Omron)
to find $D$ was in the range of 2.2 to 7 mm. The kinematic viscosity $\nu$ of
PDMS was in the range of 1000 to 10000 cS, which corresponds to the range of
the viscosity $\eta=\rho_{0}\nu$, 0.970 to 9.75 Pa$\cdot$s (the density
$\rho_{0}$ is in the range 970-975 kg/m$^{3}$). The metal disk is a
stainless-steel SUS403 of density $\rho=7.70$ g/cm$^{3}$, which can be
manipulated by a magnet placed on the cell surface. The thickness $D_{0}$
($<D$) of the disk was either 1.88, 2.87, or 3.90 mm (these three thicknesses
will be labeled as $D_{0}=2,3,$ and $4$ mm, for convenience, in Fig. 1 to 3).
We used a digital camera (EX-F1, Casio) setting the time interval in the range
of 1 to 1/60 second. The digital images were analyzed with a software, Image J.

As demonstrated in Fig. 1 (b), the velocity of falling motion of the disk
reached a constant speed, which was reproducible at a given experimental
parameter set, $\eta,$ $D_{0},$ and $h$ (an error in velocity is typically
less than 10 per cent). Here, $h$ is the thickness of two viscous liquid
films, each of which is sandwiched by an inner surface of cell plate and a
surface of the disk.

Obtaining reproducible data as in Fig. 1 (b) is highly non-trivial. The main
difficulty is to drop the disk in the cell with precisely maintaining the
relation $2h=D-D_{0}$. The falling speed is expected to be maximized when this
relation is satisfied with no tilting. Thus, we created a gate by setting two
plates of thickness comparable to $h$ at the places indicated by the two
dashed rectangles at the top of the cell in Fig. 1 (a) to help guarantee the
condition at the entry. Even with this gate, we need to carefully insert the
disk into the gate and collect the data in fast falling cases. In this way, we
could obtain reproducible data as specified above.

\begin{figure}[h]
\includegraphics[width=\textwidth]{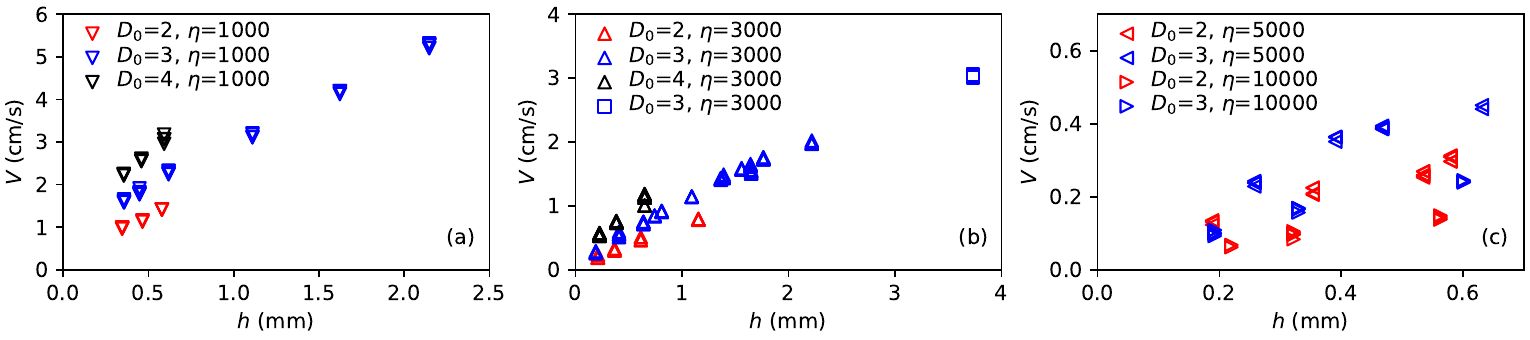} \caption{$V$ vs $h$ for $h<D_{0}%
$. (a) $\nu=1000$ cSt. (b) $\nu=3000$ cSt. (c) $\nu=5000$ and 10000 cSt.
Colors and marks differentiate $D_{0}$ and $\nu$, respectively. The labels for
$D_{0}$ and $\nu$ are given in mm and cSt. The number of the data points shown
here are 240 in total and they are obtained for 46 different parameter set
$(D_{0},\eta,h)$. (The precise values of $D_{0}$ are given in the text)}%
\label{Fig2}%
\end{figure}

In Fig. 2, we present the falling velocity $V$, obtained as a slope in the
$x-t$ plot as in Fig. 1 (b), as a function of the film thickness $h$ for
$h<D_{0}$ (except for the rightmost data set (8 data) shown by square in Fig.
2 (b), for which $h\gtrsim D_{0}$). One data point shown without error bars,
corresponds to each falling experiment. We repeated falling experiments
typically several times (in the range twice to more than 10 times) for a
single parameter set and we showed all the results in the plots. As a result,
in some case data points are closely overlapped, which suggests a small size
of errors in the measurements.

The viscosity dependence of the velocity in Fig. 2 indicates that the dynamics
is governed by gravity opposed by viscosity. In fact, the inertia can be
neglected in most (practically all, as revealed below) of the data. This is
because, writing Reynolds number Re as $\rho_{0}VL/\eta$ with introducing a
length scale $L$, the condition Re $<1$ is equivalent to $L<\eta/(\rho V)$,
where the average and minimum of $\eta/(\rho V)$ are $1.5$ and 0.02 m, while
$L$ in the present case is either $h,D_{0},D,$ or $R$, which are all smaller
than even the minimum of $\eta/(\rho V)$.

Theoretically, the change in gravitational energy per time is trivially given
as $\pi\Delta\rho gR^{2}D_{0}V$ with $\Delta\rho=\rho-\rho_{0}$. As for the
viscous dissipation, we can think of three possibilities, corresponding to
three viscous length scales, $h,$ $D,$ and $R$. The first one $\eta
(V/h)^{2}R^{2}h$, which we call "the $h$-dissipation," describes dissipation
inside the lubrication film of thickness $h$ with its volume scaling as
$R^{2}h$. The second "$D$-dissipation" $\eta(V/D)^{2}R^{2}D$ is associated
with the viscous flow developed between the cell plates separated by the
distance $D$ in a volume scaling as $R^{2}D$. The third "$R$-dissipation"
$\eta(V/R)^{2}R^{2}D$ corresponds to the dissipation with the viscous length
scale $R$ developed in a volume scaling as $R^{2}D$. If we compare the three
dissipations with their coefficients set to unity, the ratios of the $D$- and
$R$-dissipations to the $h$-dissipation are $h/D$ and $hD/R^{2}$,
respectively, both of which are less than unity in the present experiment.
More precisely, the average and median of the ratio first $h/D$ are 0.16 and
0.14, while those of the second ratio $hD/R^{2}$ are 0.039 and 0.018.

As a first trial for comparison between theory and experiment, we assume the
largest $h$-dissipation dominates over the other dissipations. Then, the
balance of the change in gravitational energy with dissipation (the
$h$-dissipation only) gives%
\begin{equation}
V\sim\Delta\rho gD_{0}h/\eta, \label{eq1a}%
\end{equation}
which can be expressed as
\begin{equation}
\eta V/(\Delta\rho gD_{0}^{2})=f(h/D_{0}) \label{eq1b}%
\end{equation}
with
\begin{equation}
f(x)\simeq x.
\end{equation}

\begin{figure}[h]
\includegraphics[width=\textwidth]{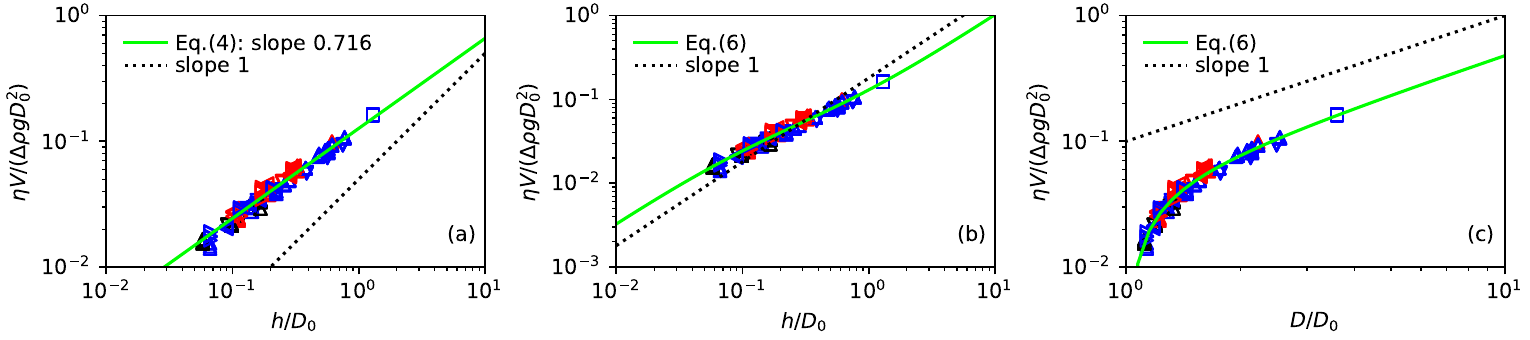}\caption{Renormalized velocity $V$
vs thickness $h$ for the 240 data in Fig. 2. (a) Fit by Eq. (\ref{eqA}) with
$\alpha=0.716$. The slope of the fit (solid line) deviates from the slope of
the dotted line, which corresponds to $\alpha=1$. (b) Successful fit by Eq.
(\ref{eq2}) with the same renormalized axes with (a). (c) The same successful
fit by Eq. (\ref{eq2}) but with a differently renormalized horizontal axis.}%
\end{figure}

In Fig. 3 (a), we plot the renormalized velocity $\eta V/(\Delta\rho
gD_{0}^{2})$ as a function of the renormalized thickness $h/D_{0}$, in the
light of Eq. (\ref{eq1b}), using all the data in Fig. 2, where the data
described by Eq. (\ref{eq1a}) should collapse onto a master curve on this
plot. This is what we observe in Fig. 3 (a): all the data in Fig. 2 (except
the dataset with $h\gtrsim D_{0}$) collapse well on a master curve.

However, the agreement is not perfect. As a result of numerical fitting, we
find that the collapsed data are well described by the following expression
\begin{equation}
\eta V/(\Delta\rho gD_{0}^{2})=k_{1}(h/D_{0})^{\alpha} \label{eqA}%
\end{equation}
with $k_{1}=0.127\pm0.001$ and $\alpha=0.716\pm0.008$ (in this fitting, the
dataset with $h\gtrsim D_{0}$ is excluded). The exponent $\alpha$ deviates
from the theoretical prediction $\alpha=1$.

This deviation cannot be explained by the slip length which has been discussed
for polymer liquids \cite{de2005soft,de1979scaling}. This is because a simple
linear extrapolation of the data of each color in Fig. 2 (a) may intersect the
horizontal axis at a value around $h=-0.1$ to $-0.3$ mm and this order of
magnitude is too large for the slip length (in the present case, it is at most
$10\mu$m).

As a second trial, we include the second largest $D$-dissipation in addition
to the $h$-dissipation, since the first trial based on the $h$-dissipation
only raised problems. Then, the new energy balance can be cast in the
following form:%
\begin{equation}
\Delta\rho gR^{2}D_{0}V=\eta V^{2}R^{2}(c_{1}/h+c_{2}/D), \label{eq4}%
\end{equation}
where $c_{1}$ and $c_{2}$ are numerical coefficients (The two energy
dissipations are added here, considering a simplest and physically plausible
case). This leads to Eq. (\ref{eq1b}) with%
\begin{equation}
f(x)=\frac{d_{1}x}{1+\frac{d_{2}x}{1+2x}}, \label{eq2}%
\end{equation}
where $d_{1}=1/c_{1}$ and $d_{2}=c_{2}/c_{1}$.

Equation (\ref{eq2}) still predicts a data collapse when $\eta V/(\Delta\rho
gD_{0}^{2})$ is plotted as a function of $h/D_{0}$, which is truly observed in
Fig. 3 (a). In Fig. 3 (b), the same plot is presented with the result of an
excellent fitting by Eq. (\ref{eq1b}) with (\ref{eq2}), where $d_{1}=$
0.34$\pm$0.01 and $d_{2}=$ 4.90$\pm$0.26. Note that even the dataset with
$h\gtrsim D_{0}$ is well fit by Eq. (\ref{eq2}) although this set is excluded
in the fitting. Since $d_{2}=c_{2}/c_{1}$ corresponds to the relative
importance of the $D$-dissipation to the $h$-dissipation as seen in Eq.
(\ref{eq4}), the present fitting predicts that the two dissipations are
comparable and the $D$-dissipation is larger than the $h$-dissipation.

However, the $D$-dissipation does not play a dominant role. This is clearly
shown in Fig. 3 (c). To understand this, note that Eq. (\ref{eq1b}) with Eq.
(\ref{eq2}) can be expressed as%
\begin{equation}
\eta V/(\Delta\rho gD_{0}^{2})=(1/c_{2})(D/D_{0})\label{eq5}%
\end{equation}
when the $D$-dissipation dominates over the $h$-dissipation, i.e., if
$c_{1}<<c_{2}$. This means if the $D$-dissipation is dominant we should
observe most of the collapsed data are on a line with slope one in Fig. 3 (c),
which is not the case. We emphasize here that the dataset with $h\gtrsim
D_{0}$ (square) in Fig. 3 (c) significantly strengthens our arguments for the
$D$-dissipation, supporting the approach to a line with slope one as $D/D_{0}$
increases. Note that thanks to the relation $D/D_{0}=1+2h/D_{0}$ when the plot
$\eta V/(\Delta\rho gD_{0}^{2})$ vs $h/D_{0}$ collapses on to a master curve,
so does the plot $\eta V/(\Delta\rho gD_{0}^{2})$ vs $D/D_{0}$.

Further insight can be obtained if we note the following limiting forms for
Eq. (\ref{eq2}):%
\begin{equation}
f(x)=\left\{
\begin{array}
[c]{ccc}%
d_{1}x &  & x<<1\\
\lbrack d_{1}/(1+d_{2}/2)]x &  & x>>1
\end{array}
\right.  \label{eq3}%
\end{equation}
These limiting behaviors can be observed in Fig. 3 (b) with the aide of the
dotted line with slope one. In between there exists a quasi-straight-line
region with data collapsed. Such a straight-line region, if exists, should
possesses a slope smaller one, since the first coefficient $d_{1}$ is larger
than the second coefficient $d_{1}/(1+d_{2}/2)$ in Eq. (\ref{eq3}) and, thus,
the second slope-one region ($x>>1$) is shifted below the first ($x<<1$), as
can be confirmed in Fig. 3 (b). This is the reason we obtained $\alpha$ less
than one in the above. Note that the dataset with $h\gtrsim D_{0}$ is
consistent with this argument: all the data in Fig. 3 (b) are well on the fit
in the region between the apparent scaling with slope $\alpha$ and the linear
scaling with slope one for $x>>1$ in Eq. (\ref{eq3}), where the set with
$h\gtrsim D_{0}$ is in the region rather close to the starting of the linear
scaling. Further note that while the linear scaling for $x<<1$ corresponds to
the real scaling for the $h$-dissipation, that for $x>>1$ does not correspond
to the real scaling for the $D$-dissipation.

For completeness, we explain the reason we can neglect the $R$-dissipation,
although this dissipation is expected to be the smallest. We rewrite the
balance in Eq. (\ref{eq4}) with including the $R$-dissipation: $\Delta\rho
gR^{2}D_{0}V=\eta V^{2}R^{2}(c_{1}/h+c_{2}/D+c_{3}D/R^{2}),$ to obtain Eq.
(\ref{eq1b}) with%
\begin{equation}
f(x)=\frac{d_{1}x}{1+\frac{d_{2}x}{1+2x}+d_{3}\left(  \frac{D_{0}}{R}\right)
^{2}x(1+2x)}. \label{eq7}%
\end{equation}
This form predicts that, if $d_{3}$ term cannot be neglected, the collapse
onto a master curve cannot be observed in Fig. 3 (because the data includes
those obtained for different $D_{0}$ although $R$ is fixed), which is not
true. Thus, we can judge that the remaining $R$-dissipation can be safely
neglected from our argument to explain the collapsed data.

We may learn two lessons from the present analysis. (1) When a log-log plot is
seen in a limited range, in principle, it tends to seem a straight line, and
when the slope is fairly close to the one predicted from a simple dimensional
analysis, one tends to overlook other important physical origin of the
problem. (2) In such a case, or even when such an apparent scaling regime
spans over a few or several decades, when the exponent deviates from the one
predicted from a simple argument, it is worth considering competition of a few
scaling regimes, as simply demonstrated in the above.

The present apparent scaling regime demonstrated in Fig. 3 (a) spans at least
over one order of magnitude, but a limited range. According to the theory, our
data happen to correspond to the ones in the crossover region in between the
two asymptotic regimes defined in Eq. (\ref{eq3}). Judging from the behavior
of Eq. (\ref{eq3}) shown in Fig. 3 (b), we need to further explore regions
$h/D_{0}<0.01$ and $>1$ to completely confirm our argument by explicitly
establishing real scaling regimes originating from $h$- and $D$-dissipations
(in addition to the apparent scaling regime). This implies nearly over four
orders of magnitude for $h/D_{0}$ in total, which is an important but
experimentally challenging future problem. Note here that the real scaling for
$D$-dissipation is already reasonably well confirmed in Fig. 3 (c) as
mentioned above with the help of the dataset with $h\gtrsim D_{0}$.

We discuss here a semi-quantitative estimate for $c_{1}=1/d_{1}$, which can be
obtained in the following way. The velocity gradient could be precisely $\eta
V/h$ (with assuming a simple Couette shear flow as in the film between two
infinite plates separated by $h$) almost everywhere on the disk surface except
near the edge in the small $h$ limit. Then, in this limit, the total drag
force might be obtained simply by multiplying this gradient by the area of
both disk surfaces $2\pi R^{2}$. Balancing the force thus obtained with the
gravitational force $\pi R^{2}D_{0}\Delta\rho g$, we could obtain a plausible
value $d_{1}=1/2$ (this leads to a one-parameter fitting, resulting in
$d_{2}=9.25\pm0.08$). It is interesting this plausible value is not far from
the previous value ($d_{1}=0.34$). However, the data for small $h/D_{0}$ in
the present study slightly better support the previous fitting. This issue
could be another motivation for a separate study for $h/D_{0}<0.01$ and $>1$.

As for the coefficient $c_{2}$, when not a solid disk but a fluid drop
(viscosity $\eta^{\prime}$) moving in a Hele-Shaw cell filled with another
viscous fluid (viscosity $\eta$), the formula $V=(1/12)\Delta\rho gD^{2}/\eta$
is derived in the limit $\eta^{\prime}=0$ and $D=D_{0}$ in
\cite{Maxworthy1986} from a result in \cite{TAYLORSAFFMAN1959}, which has been
confirmed different methods \cite{bush1997anomalous,gallaire2014marangoni}.
While the present case corresponds $\eta^{\prime}>>\eta$ and $D=D_{0}+2h$, if
the formula could be applied, it would predict $c_{2}=12$. This estimate is
also not far from the value of $c_{2}$ obtained from the two-parameter fitting
as well as the one-parameter fitting: in the former case, $c_{2}=d_{2}%
/d_{1}=4.9/0.34$ and in the latter case $c_{2}=9.25/(1/2)$.

The present result can be appreciated in the context of a Stokes drag friction
$F$ for a solid disk moving in a direction perpendicular to the axis, i.e., in
a direction of the disk plane, in a confined geometry specified by two
parameters: a measure of confinement $C=R/D$ and the aspect ratio of the disk
$A=D_{0}/R$. Under no confinement ($C=0$) for a disk of zero thickness
($A=0$), the drag friction was given by Oberbeck \cite{Oberbeck} as
$F_{O}=32\eta VR/3$. After the case of $A=0$ and small $C$ was studied
\cite{brenner1962effect,davis1991slow}, the case of $C=0$ and $A\neq0$ was
considered \cite{trahan1999velocity}. Then, the case of $C\neq0$ and $A\neq0$
was reported \cite{trahan2006stokes}, where the case with $h<$ $D_{0}$ was not
studied, which is the main focus of the present study. In terms of the drag
friction force $F$ $(=KF_{O})$, the present study ($0.05\lesssim
h/D\lesssim0.3$) predicts the following viscous drag and the drag coefficient:
$F=\pi\eta VR^{2}(c_{1}/h+c_{2}/D)$ and $K=(3\pi R/32)(c_{1}/h+c_{2}/D)$ with
$c_{1}$ and $c_{2}$ given through the above values of $d_{1}$ and $d_{2}$.
Note that $K$ represents the enhancement of the friction due to confinement
and thus becomes significantly larger than one under a strong confinement in
which $R\gg h$ and $D$. This implies that the falling velocity in the present
confined case is $(3\pi/4)(D_{0}/R)/K$ times slower than the non-confined case.

The present falling disk problem is inherently different from the case in
which the disk is replaced with a fluid "disk." Such a case is studied under
no existence of surfactants \cite{yahashi2016}, in which, the thickness of
lubricating film $h$ and the disk thickness (or, the shape of fluid drop) are
dynamically determined and thus dependent on $V$, with the former governed by
the law of Landau, Levich, and Derjaguin (LLD)
\cite{LandauLevich,Derjaguin1943}. The drag force $F$ thus found scaled with
$V^{1/3}$, in contrast to the present case, in which $h$ is fixed by the
(fixed) disk thickness and $F$ scales simply with $V$.

More complex cases with rigid and mobile surfactants are relevant if fluid
drops of various kind, such as armored bubbles or drops, gelified drops,
bubbles and drops with surfactants, are utilized. Such cases have been
explored mainly on a smaller length scale appropriate for microfluidics
\cite{huerre2014bubbles,cantat2013liquid}. While direct comparison of such
studies with the present case may be irrelevant because of the inherent
difference in how the thickness of the lubricating film is determined, the
study of cases with surfactants on a larger cm scale as in the present study
will be an important future problem.

In the present study, we investigated the falling velocity of a solid disk in
viscous liquid in a confined space and the drag friction acting on the disk.
We identified an apparent scaling regime, in which two scaling regimes
cooperating. The present study provides a simple possibility of explaining an
apparent scaling law, which would be useful to any physical scientists,
considering the generality and strength of scaling analysis in science. In
addition, fundamental understandings on fluid flow at low Reynolds numbers
provided in the present study could be valid for small objects in less viscous
fluid such as water. Accordingly, the present study are relevant to various
fundamental issues and applications, for example, in microfluidics,
bioconvection and active matter, in which viscous friction acting on small
objects are highly important.

\begin{acknowledgments}
This work was partly supported by JSPS \ KAKENHI Grant Number JP19H01859.
\end{acknowledgments}

\end{document}